\documentclass[pdflatex,sn-mathphys-num]{sn-jnl}

\usepackage{graphicx}%
\usepackage{multirow}%
\usepackage{amsmath,amssymb,amsfonts}%
\usepackage{amsthm}%
\usepackage{mathrsfs}%
\usepackage[title]{appendix}%
\usepackage{xcolor}%
\usepackage{textcomp}%
\usepackage{manyfoot}%
\usepackage{booktabs}%
\usepackage{algorithm}%
\usepackage{algorithmicx}%
\usepackage{algpseudocode}%
\usepackage{listings}%
\usepackage{geometry}
\usepackage{multicol}
\newgeometry{textwidth=180mm,textheight=243mm}
\usepackage{placeins}
\usepackage{float}

\theoremstyle{thmstyleone}%
\theoremstyle{thmstyletwo}%

\theoremstyle{thmstylethree}%

\begin{document}

\title[Article Title]{Vacuum Ultraviolet Dual-Comb Spectroscopy}

\author[1]{\fnm{John J.} \sur{McCauley}}
\equalcont{These authors contributed equally to this work.}

\author[1,2]{\fnm{Dylan P.} \sur{Tooley}}
\equalcont{These authors contributed equally to this work.}

\author*[1]{\fnm{R. Jason} \sur{Jones}}\email{rjjones@arizona.edu}

\affil[1]{\orgdiv{James C. Wyant College of Optical Sciences}, \orgname{University of Arizona}, \orgaddress{\street{}\city{Tucson}, \state{AZ} \postcode{85721}, \country{USA}}}

\affil[2]{\orgdiv{Department of Physics}, \orgname{University of Arizona}, \orgaddress{\street{}\city{Tucson},  \state{AZ} \postcode{85721}, \country{USA}}}


\abstract{\unboldmath The optical frequency comb has made a significant impact in precision spectroscopy and on our ability to probe atomic, molecular and,
recently, nuclear transitions to further our understanding of their fundamental properties and how their dynamics and complex interactions affect the observed world. To expand the energy scales and types of systems that can be studied, frequency comb sources from terahertz to vacuum ultraviolet frequencies and beyond have been pursued. 
Dual-comb spectroscopy, enabled by the development of these frequency comb sources, allows broadband absorption measurements of complicated spectra,  
exceeding the limitations of direct, single-comb spectroscopy. To date, however, the dual-comb approach has not been able to directly access many important transitions that lie at challenging vacuum ultraviolet wavelengths. Here, we demonstrate dual-comb spectroscopy in the vacuum ultraviolet utilizing intracavity high harmonic generation. This multi-harmonic source is used to measure molecular absorbance spectra at $\lambda=210$~nm and $\lambda=149$~nm from room-temperature samples of acetylene and ammonia, respectively. These measurements resolve the Doppler broadened structure of congested molecular spectra with absolute frequency accuracy. Noise contributions to the vacuum ultraviolet dual-comb spectroscopy measurements are characterized, guiding future efforts and technological development in this region.} 



\maketitle
\vspace{-10mm}
\begin{multicols}{2}


The ultraviolet region of the electromagnetic spectrum contains a large number of strong atomic and molecular transitions originating directly from ground states that provide critical quantitative information needed for understanding fundamental processes and facilitating advances in science and technology. Although much progress has been made extending precision laser sources to the ultraviolet, the vacuum ultraviolet (VUV) ($10<\lambda<200$~nm)---where atmospheric gases such as  O$_2$ strongly absorb~\cite{international2007iso}---has proved technically challenging for source generation and spectrometer design. Advances in high-resolution VUV spectroscopy impact a variety of industries and research fields. For example, spectroscopy of strong ultraviolet transitions in neutral and low ionization stages of atomic and molecular populations can greatly aid plasma diagnostics in many areas including monitoring of critical species in fusion reactors and degradation of plasma-facing materials~\cite{bechu2020direct}, characterization of shock layer boundaries in hypersonics~\cite{holloway2022sensitivity,krish2022spectrally}, and studying effects of surface-plasma interactions on number densities of reactive species for medicine, surface decontamination~\cite{peverall2019spectroscopy}, and semiconductor etching processes~\cite{hori2006progress}, to name a few. Testing of fundamental physical laws, constants, and benchmarking quantum electrodynamics often rely on precision measurements of VUV transitions in simple molecules (e.g. H$_2$, D$_2$) and direct measurements of Rydberg states~\cite{hussels2022improved,roth2023high}, multiply-charged ions~\cite{berengut2011electron}, and even nuclear transitions~\cite{zhang2024frequency}. Additionally, exoplanet research requires accurate and high-resolution VUV absorption cross-section data of molecular gases such as carbon dioxide, acetylene, and ammonia to inform models of the photochemistry induced by stellar radiation in the upper-atmospheres of these planets~\cite{venot2013high,ranjan2020photochemistry,fleury2025high,pratt2023high}. 

While several methods for VUV spectroscopy exist, each has its own limitations and complexities in addressing these applications. 
Dispersive spectrometers are suitable for many broadband studies~\cite{trichard2017evaluation}, but require complex calibration for accurate measurements~\cite{FROHLERBACHUS2021107427}, and are ultimately limited in their spectral resolution~\cite{samson2000vacuum}. VUV Fourier-transform spectrometers (FTS) have emerged offering higher resolution and simplified calibration for emission spectroscopy~\cite{thorne1987fourier,griesmann1999nist} and isolated examples of absorption spectroscopy with a discharge light source~\cite{murray1994vacuum}. However, for VUV FTS absorption spectroscopy, a synchrotron beamline is often required for a coherent continuum source~\cite{yoshino1995combination,matsui2003high,deOliveira2011high}. 

\begin{figure*}[b]
\centering
\includegraphics[scale=0.999]{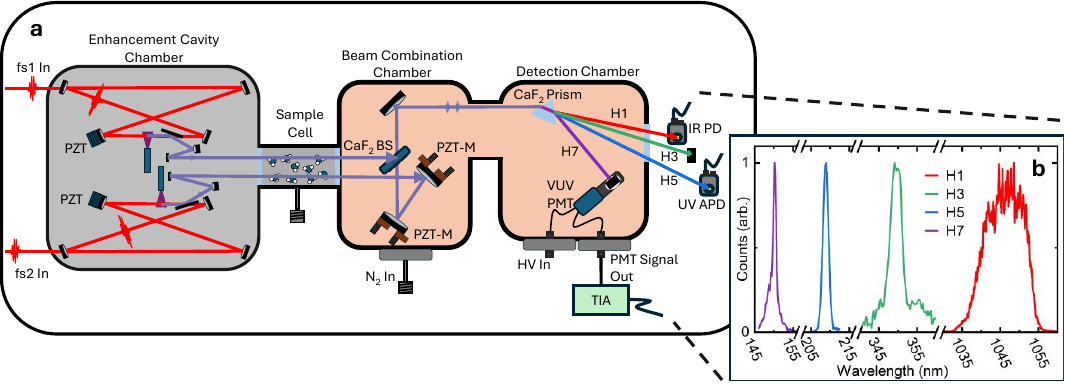}
\caption{\textbf{Intracavity High Harmonic Generation and Dual-Comb Spectroscopy Beamline.} \textbf{a}, Experimental schematic of intracavity high-harmonic generation showing each comb's out-coupled collinear harmonics aligned through a sample cell and into a nitrogen-purged chamber where they are overlapped on an ultraviolet beam splitter (BS) with piezo-controlled mirrors (PZT-M). Finally, the harmonics are separated with a CaF$_2$ prism, and the fundamental (H1) is measured on an infrared photodiode (IR PD), the 5$^\text{th}$ harmonic (H5) on an ultraviolet-enhanced avalanche photodiode (UV APD), and the 7$^\text{th}$ (H7) on a solar-blind photomultiplier tube (PMT). Vacuum chamber feed-throughs carry the high voltage (HV) bias to the PMT and output photocurrent from the PMT to an external trans-impedance amplifier (TIA). \textbf{b}, Low-resolution measurements of fundamental and harmonic spectra detected on various spectrometers.}\label{fig_Experimental_Spectra}
\end{figure*}

Twenty years ago, the first demonstrations of intracavity high harmonic generation (iHHG) extended optical frequency combs to the vacuum ultraviolet~\cite{jones2005phase,gohle2005frequency}. Since then, many experiments have used iHHG to perform direct frequency comb spectroscopy in the VUV with unrivaled spectral resolution and absolute frequency calibration (see, for example ~\cite{ozawa2008high,yost2009vacuum,bernhardt2012vacuum,ozawa2013vuv,benko2014extreme}). Recently, high-precision measurements of the thorium nuclear isomer have showcased the impact of the VUV frequency comb~\cite{zhang2024frequency}. With the possibility of a nuclear clock based on this transition~\cite{elwell2024laser,tiedau2024laser} comes the exciting prospect of testing our physical models through measurements of fundamental physical constants, and highlights the importance of precision VUV laser spectroscopy and source development~\cite{uzan2011varying,Flambaum2016LorentzInvariance,safronova2018search}.

However, a continuing limitation of VUV direct frequency comb spectroscopy is the difficulty, with only a single comb, of measuring multiple absorption features across the comb's broad bandwidth. 
This has limited direct comb spectroscopy in the VUV to narrow-band measurements of only sparse and well isolated transitions. To overcome this challenge at longer wavelengths, dual-comb spectroscopy (DCS) has proved to be a powerful method for broadband measurements of congested atomic and molecular spectra with resolution ultimately limited only by the linewidth of the individual comb teeth~\cite{Coddington:16,picque2019frequency}. In DCS, pulses from a pair of frequency combs with slightly different repetition frequencies ($\Delta f_{rep}=f_{rep,1}-f_{rep,2}$) are overlapped, building up an interferogram (IGM) as successive pulses step through each other. The Fourier transform of this IGM yields an optical spectrum down-converted to radio frequencies. DCS has many of the same strengths as FTS, such as high resolution and broad bandwidth, with the additional simplicity of no moving parts and the capability for more rapid measurements of time-resolved events~\cite{bergevin2018dual,zhang2021burst}. Further, DCS provides absolute frequency measurements with high accuracy due to the coherent structure of the two optical frequency combs. However, DCS has previously been limited to longer wavelengths, only recently reaching the deep ultraviolet (DUV) ($200<\lambda<280$~nm)~\cite{mccauley2024dual,hofer2025free,li2025deep,chang2024multi,furst2024broadband,muraviev2024dual,xu2024near}. Here, we use iHHG of two ytterbium-based infrared combs to generate the DUV 5$^\text{th}$ harmonic ($\lambda=210$~nm) and VUV $7^\text{th}$ harmonic ($\lambda=150$~nm) for DCS. At both harmonics, we demonstrate the capabilities of the DCS system with absolute frequency measurements of Doppler-resolved absorbance spectra of room-temperature molecular gases relevant to exoplanet photochemistry studies. This work extends the coverage of DCS to a previously inaccessible spectral regime, enabling fast and absolute frequency measurement capabilities with  high spectral-resolution and  bandwidth in the VUV.

\section{Results}\label{sec2}

\subsection{Intracavity High Harmonic Dual-Comb Spectrometer}\label{sec3}
The experimental schematic is shown in Fig.~\ref{fig_Experimental_Spectra}. A pair of home-built ytterbium fiber lasers generate near identical optical pulse trains with center wavelengths near $\lambda=1050$~nm, repetition frequencies $f_{rep}=77.8$~MHz, pulse durations of 150~fs, and average powers of 20~W. (see Methods for more details). These pulse trains are independently coupled into passive femtosecond enhancement cavities (fsECs) with similar cavity lengths, where intracavity average powers reach 5~kW~\cite{jones2000stabilization,jones2002femtosecond}. The Pound-Drever-Hall (PDH) technique is used to lock the length of each fsEC to the incident pulse train using fast piezo-electric transducers (PZTs) on fsEC mirrors.   
Xenon gas jets are introduced after the cavity foci, where peak intensities exceed \mbox{$5\times10^{13}~\text{W}/\text{cm}^2$}. High harmonic generation in the Xe gas generates odd harmonics of the fundamental, which are out-coupled with sapphire grazing incident plates (GIPs) at 70$^\circ$ angle of incidence. These GIPs have a high Fresnel reflectance across a broad bandwidth of s-polarized 
ultraviolet light with an anti-reflection coating for the fundamental beam ensuring low loss for the pulses circulating in the fsECs~\cite{fischer2022efficient}. After the GIPs, the out-coupled harmonics are collimated by a curved mirror. 

To achieve spatial overlap of the generated collinear harmonic pulse trains, we steer the beams through a CaF$_2$-windowed sample cell and into a nitrogen purged chamber, using Al+MgF$_2$ coated mirrors, and combine them on a coated CaF$_2$ VUV beamsplitter. Piezo-actuated mirror mounts allow for fine tuning of the VUV spatial overlap while preserving the nitrogen purged environment. The harmonic orders of the combined beams are then spatially separated by a CaF$_2$ prism. The $7^\text{th}$ harmonic VUV combs 
are detected using a solar-blind 
photomultiplier tube (PMT). The fundamental, 3$^{\text{rd}}$, and 5$^{\text{th}}$ harmonics 
are picked off and aligned out of the combination chamber. The fundamental dual-comb IGMs are used for timing and triggering during data acquisition and the 5$^{\text{th}}$ is detected on a UV-extended silicon avalanche photodiode (APD). From the measured PMT photocurrent, we estimate the 7$^\text{th}$ harmonic power per comb on the detector to be $\sim$0.6~$\mu$W, sufficient for dual-comb IGM detection. Taking into account the expected reflection and transmission losses from the optical elements in the VUV beamline, we conservatively estimate our out-coupled 7$^\text{th}$ harmonic VUV power to be $\sim$11.5$~\mu$W per comb. The low-resolution harmonic spectra, as shown in Fig.~\ref{fig_Experimental_Spectra}b, are measured on various spectrometers. The current dual-comb source extends to the VUV 7$^\text{th}$ harmonic; however, we expect the iHHG process is also producing significant powers of extreme ultraviolet harmonics ($\lambda<120$~nm) which are currently absorbed by use of CaF$_2$ transmissive optics. 

\subsection{Dual-Comb Characterization}\label{sec3}
\begin{figure}[H]
\centering
\includegraphics[]{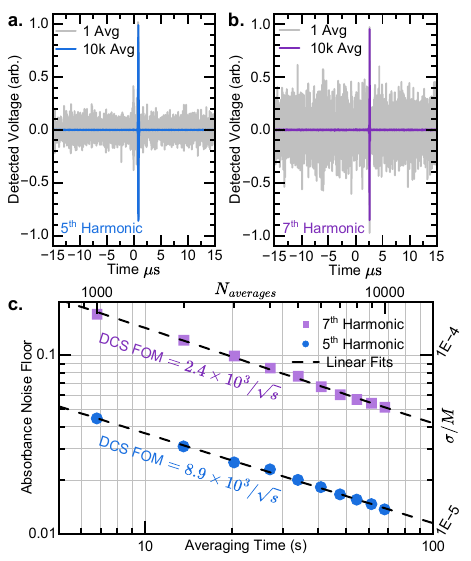}
\caption{\textbf{Dual-Comb Interferogram Detection and Averaging} \textbf{a-b}, Detected 5$^{\text{th}}$ and 7$^{\text{th}}$ harmonic single-shot interferograms (gray) with centerbursts arbitrarily scaled to $\pm1$~unit are phase-corrected and averaged 10,000 times (\textbf{a.} blue, \textbf{b.} purple). \textbf{c}, For averaging time $\tau$, the noise floor of the absorbance spectra decreases with $\sqrt{\tau}$. Here, both 5$^{\text{th}}$ and 7$^{\text{th}}$ harmonic interferograms are processed to resolve $M=1000$ spectral elements across their respective optical bandwidths, giving $\Delta\nu_{res}=4$ and 7~GHz, respectively.}\label{fig_IGM_Averaging}
\end{figure}

With an evacuated sample cell, IGMs from the 5$^\text{th}$ harmonic APD and 7$^\text{th}$ harmonic PMT are recorded on a 14-bit digitizer with $\Delta f_{rep}=147$~Hz. We average the IGMs in time by measuring and correcting the envelope position and slow phase variation of the center-burst. In past UV studies, coherent averaging has been effective even with photon-limited power levels~\cite{xu2024near}. Here, the resonant conditions of the fsECs constrain the arbitrary selection of the combs' carrier-envelope-offset frequencies ($f_{ceo}$), a degree-of-freedom typically required for coherent averaging. Therefore, we must phase-correct IGMs individually before averaging, requiring the single-shot time-domain signal-to-noise ratio (SNR) of IGMs to be $>1$. Fig.~\ref{fig_IGM_Averaging} shows representative single-shot IGMs and averages of 10,000 records.

\begin{figure*}[b]
\centering
\includegraphics[]{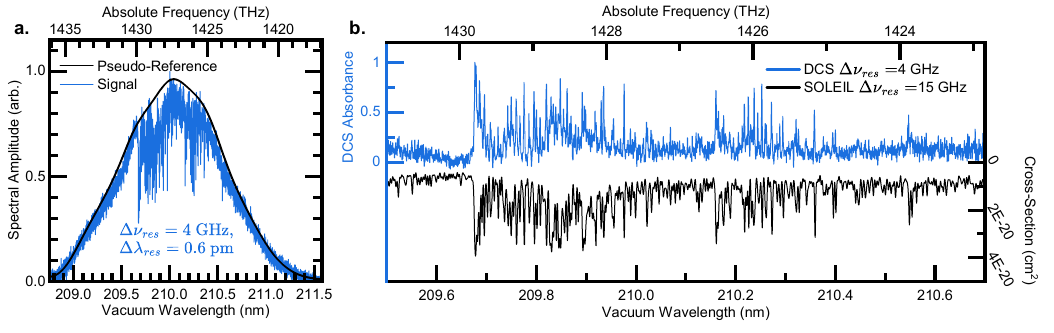}
\caption{\textbf{Deep ultraviolet dual-comb spectroscopy of acetylene with the 5$^{\text{th}}$ harmonic at $\lambda=210$~nm}. \textbf{a}, The dual-comb transmission spectrum processed with a resolution of 4~GHz and 10,000 averages (blue) and a smoothed pseudo-reference (black). \textbf{b}, The dual-comb absorbance spectrum (blue) is compared to the absorption cross-section of acetylene (black) measured at the SOLEIL synchrotron with a resolution of 15~GHz~\cite{fleury2025high}.}\label{fig_5th_C2H2}
\end{figure*}

For each harmonic, the averaged IGM is apodized by a Blackman-Harris window selected to achieve an intended instrument resolution. Here, the 5$^\text{th}$ (7$^\text{th}$) harmonic has a resolution of 4~GHz (7~GHz) across its 4~THz (7~THz) bandwidth, resulting in $M=1000$ frequency resolved elements in each harmonic spectrum. While the best resolution attainable from the full length, averaged IGM is ultimately limited only by the comb tooth spacing, $f_{rep}$, in these measurements we target resolving power that will be physically relevant to resolve Doppler-broadened molecular absorption features in measurements reported in Sections~\ref{subsec_5_C2H2} and \ref{subsec_7_NH3}. 
The magnitude of the apodized IGM's Fourier transform gives a transmitted dual-comb power spectrum, which is normalized to a featureless absorbance spectrum and used to analyze the performance of the DUV and VUV DCS absorbance measurement. For 10,000 averages recorded in $\tau\sim$1 minute, the standard deviation of the absorbance noise floor across the transmission spectrum's full-width at half-maximum reaches $\sigma=$0.015 for the 5$^{\text{th}}$ harmonic and $\sigma=$0.05 for the 7$^{\text{th}}$. As shown in Fig.~\ref{fig_IGM_Averaging}, the noise floor approaches these values with the expected $\sqrt{\tau}$ dependence~\cite{newbury2010sensitivity}. From this performance, we calculate a DCS figure of merit as FOM$=\frac{M}{\sigma\sqrt{\tau}}=8.9\times10^3/\sqrt{\text{s}}$ for the 5$^{\text{th}}$ harmonic and FOM$=2.4\times10^3/\sqrt{\text{s}}$ for the 7$^{\text{th}}$. These results demonstrate that our DUV and VUV dual-comb spectrometer is capable of measuring molecular absorption with low minimum detectable absorbance and high signal to noise, with resolution sufficient for resolving typical molecular Doppler widths at room temperature.

\subsection{5$^\text{th}$ Harmonic Acetelyne Absorbance}\label{subsec_5_C2H2}
Using the 5$^\text{th}$ harmonic of the DCS laser system, we target an absorption feature near $\lambda=210$~nm in the ${\tilde{A}\leftarrow\tilde{X}}$ band of acetylene (C$_2$H$_2$)~\cite{fleury2025high,BENILAN2000C2H2} by tuning the center of each comb's infrared spectra to near $\lambda=1050$~nm. We  introduce a pressure $P=50$~Torr of acetylene gas into a sample cell with length $L\sim14$~cm. For this measurement, $\Delta f_{rep}=101$~Hz. 10,000 IGMs are recorded, phase-corrected, averaged, and apodized to produce a single IGM. The retrieved dual-comb transmission spectrum ($S_{sig}$) is shown in Fig.~\ref{fig_5th_C2H2}a with a resolution of 4~GHz (0.6~pm) as determined by the apodization window. 

The retrieved transmission spectrum 
is composed of radio-frequency (RF) carriers below $f_{rep}/2=39$~MHz. However, because the mutually-coherent combs are referenced to a continuous-wave (CW) laser of known optical frequency (see Methods), this RF transmission spectrum is self-calibrated to absolute optical frequency. The transmission spectrum shows many congested absorption features, and a pseudo-reference ($S_{ref}$) is constructed to match the background spectral shape. Finally, we calculate the absorbance spectrum $A=\ln(S_{ref}/S_{sig})$ shown in Fig.~\ref{fig_5th_C2H2}b. To validate our absolute frequency measurement of this acetylene absorption band, we compare to the acetylene cross-section measured with a resolution of 15~GHz on the Fourier-transform spectrometer DESIRS beamline at the SOLEIL synchrotron~\cite{fleury2025high}. We observe excellent agreement in the independent wavelength calibration of the absorption spectra as shown in Fig.~\ref{fig_5th_C2H2}. Additionally,  the previously reported cross-section of $4\times 10^{-20}~\text{cm}^2$ for the molecular band head near $\lambda=209.7$~nm would predict an absorbance of $A=1$, given the acetylene cell temperature and pressure, which is very near the measured absorbance.

\subsection{7$^\text{th}$ Harmonic Ammonia Absorbance}\label{subsec_7_NH3}
\begin{figure*}[]
\centering
\includegraphics[]{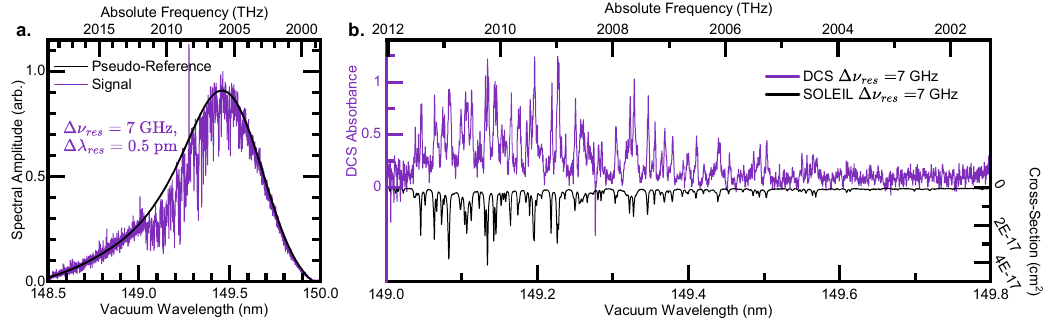}
\caption{\textbf{Vacuum ultraviolet dual-comb spectroscopy of ammonia with the 7$^{\text{th}}$ harmonic at $\lambda=149$~nm}. \textbf{a}, The dual-comb transmission spectrum processed with a resolution of 7~GHz and 15,000 averages (purple) and a smoothed pseudo-reference (black). \textbf{b}, The dual-comb absorbance spectrum (purple) is compared to the absorption cross-section of ammonia (black) measured at the SOLEIL synchrotron with a resolution of 7~GHz~\cite{pratt2023high}.}\label{fig_7th_NH3}
\end{figure*}

Next, we use the VUV 7$^\text{th}$ harmonic to target an ammonia (NH$_3$) absorption feature in the ${\tilde{B}\leftarrow\tilde{X}}$ band centered at $\lambda=149$~nm. From the previous 5$^{th}$ harmonic acetylene measurement, we tune our infrared comb spectra by $\Delta\lambda=7$~nm, which tunes the VUV spectra of the 7$^\text{th}$ harmonic by 1~nm (13.5~THz) to overlap with this absorption feature. This adjustment is greater than the VUV bandwidth in any single position and demonstrates the wavelength agility of this approach for a targeted absorption study. A previously evacuated sample cell with length $L\sim4.5$~cm is filled with a pressure $P=0.8$~Torr of ammonia. For this measurement, $\Delta f_{rep}=188$~Hz. With processing methods similar to the 5$^{\text{th}}$ harmonic absorbance study, 15,000 recorded 7$^{\text{th}}$ harmonic VUV IGMs are phase-corrected, averaged, and apodized to yield an absolute frequency transmission spectrum with a resolution of 7~GHz (0.5 pm) as shown in Fig.~\ref{fig_7th_NH3}a. The absorbance spectrum, as shown in Fig. ~\ref{fig_7th_NH3}b, is again obtained from the transmission spectrum and a constructed pseudo-reference. A comparison of this first DCS VUV absorbance spectrum to the previously reported cross-section~\cite{pratt2023high} demonstrates the quality of DCS in the VUV and its ability to resolve complex molecular absorption bands with absolute frequency calibration.

\section{Discussion}\label{sec_discussion}

\begin{figure}[H]
\centering
\includegraphics[]{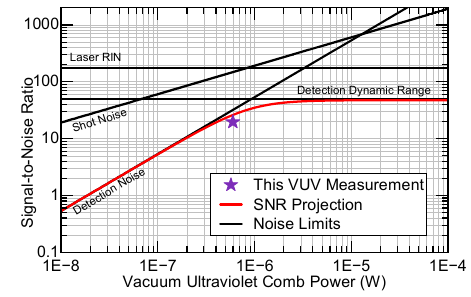}
\caption{\textbf{Vacuum ultraviolet dual-comb spectroscopy noise analysis.} For $M=1000$ frequency-resolved elements ($\Delta\nu_{res}=7$~GHz resolution across $\Delta\nu_{BW}=7$~THz optical bandwidth), $\tau=68$~s averaging time, and 0.6~$\mu$W power on detector per comb, our vacuum ultraviolet dual-comb measurement has an SNR near 20 (star). Projected SNR at other detected comb powers (red) is calculated from various noise source contributions (black).}\label{fig_NoiseProjection}
\end{figure}

While the DUV and VUV dual-comb spectrometer is already capable of resolving the absorption structure of room-temperature molecular gases, future developments will allow for improvements in the combination of SNR, resolution, and averaging times, which is characterized in the DCS FOM. To better understand the noise currently dominating our measurement at the 7$^{\text{th}}$ harmonic, we calculate expected contributions from detection noise, shot noise, laser relative intensity noise (RIN), and dynamic range~\cite{newbury2010sensitivity}. See Methods for details of these noise contribution calculations. Fig.~\ref{fig_NoiseProjection} shows the expected SNR at various detected VUV comb powers (red) along with the SNR and VUV power of this initial demonstration (star). This measurement is dominated by detection noise and detection dynamic range. 
\par
The detection requirements of DCS are uniquely challenging for the VUV. In particular, a high sensitivity, low noise, high bandwidth detector is necessary. Here, we use a solar-blind PMT whose specifications do meet these requirements; however, the low PMT photocurrent is particularly susceptible to stray current noise pickup before it is converted to a more robust voltage signal by a sufficiently fast trans-impedance amplifier (TIA). We suspect that the stray current noise picked up in the vacuum chamber feed-through between the VUV PMT and external TIA is the main contributor to the detection noise limit and calculate its contribution, shown in Fig.~\ref{fig_NoiseProjection}. Additionally, the ratio of the PMT's maximum output current to this stray noise sets the detection dynamic range limit. In future experiments, more robust feed-through design and/or a fast, low noise, vacuum compatible TIA will improve these limits from the detection noise and the detection dynamic range. The latter of these will be further benefited by a higher maximum output current PMT. With these improvements, we estimate that the dual-comb FOM could increase by nearly an order of magnitude, yielding improvements in the SNR, resolution, and/or averaging time compared to measurements presented here. Alternatively, 
the quality of VUV DCS measurements would significantly improve with developments in the VUV detectors available with high bandwidth, high responsivity, and vacuum compatibility, such as VUV APDs.

With these detection improvements, increasing the detected VUV power will  increase the dual-comb FOM until reaching a limit from laser RIN. This RIN limit is conservatively estimated in Fig.~\ref{fig_NoiseProjection} where shot noise contributions likely obscure the true RIN contribution. In the case of a RIN-limited measurement, VUV balanced detection would benefit performance further. We expect future measurements with increased VUV power on detector due to more efficient harmonic separation and other VUV beamline improvements. While out-coupled 7$^\text{th}$ harmonic VUV power per comb exceeding 10~$\mu$W is sufficient for the measurements presented here, in other measurements we out-coupled over 20~$\mu$W per 7$^\text{th}$ harmonic VUV comb by increasing xenon backing pressure. However, this higher VUV power introduces additional difficulties for long acquisition periods and is not particularly impactful before first addressing the above detection limitations.

These results demonstrate DCS in the VUV for the first time. We resolve the Doppler-broadened line widths of complex molecular absorbance spectra with high SNR. The absolute frequency and photometric accuracy of the DCS system will allow for further targeted absorption cross-section studies in the VUV with modest table-top infrastructure. Additionally, iHHG dual-comb sources provide a scalable approach for extending DCS even farther into the extreme ultraviolet. This regime, beyond the transmission band of  CaF$_2$ optics, can make use of previously demonstrated wavefront-divsion beamsplitters and Fresnel zone plate optics~\cite{samson2000vacuum,benko2014extreme} to overlap and record IGMs of higher harmonics. 
The establishment of DCS in the VUV provides a powerful spectroscopic method in this challenging spectral regime, enabling fast acquisition, absolute frequency accuracy, broad bandwidth, and spectral resolution limited only by the linewidth of the comb teeth. The ability to precisely probe complex atomic and molecular quantum structure and dynamics through  excitation of high energy transitions directly from the ground state opens new opportunities for quantitative measurements that can have a dramatic impact in fields ranging from  plasma diagnostics to astrophysics and fundamental science.

\bibliography{sn-bibliography}

\newpage
\section{Methods}\label{sec5}
\subsection{Infrared Comb Source}

Each home-built frequency comb originates from a nonlinear-polarization-rotation mode-locked ytterbium-doped fiber oscillator whose output is amplified in a dual-clad fiber preamplifier. The average power output of the amplifier is nominally $1-2$~W. 
The pulses are stretched to $\sim2$~ps pulse duration and spectrally filtered using a Martinez-stretcher~\cite{martinez1987stretcher}, providing control over the portion of the spectrum to be amplified by the subsequent photonic crystal fiber power amplifier (PCF). The commercial PCF module generates up to $50$~W of average power, although we have only used outputs near to $20$~W in the current experiments. The pulses are compressed to durations of $\sim150$~fs and coupled into the fsEC. We estimate our enhancement factor to be $250$ giving $\sim5$~kW average intracavity power, with a beam diameter of approximately $\sim$22~$\mu$m at the cavity focus. Shortly after the cavity focus we introduce a xenon gas jet where the peak intensity exceeds $5\times10^{13}$~W/cm${^2}$, which is sufficient to drive the high harmonic process. The gas jet originates from a glass capillary with inner diameter of 100~$\mu$m with a xenon backing pressure near 1500~Torr. The gas jets are on piezo-driven translation stages to achieve optimal harmonic yield.  

We establish the mutual coherence necessary for DCS measurements between these two combs with optical phase locks of comb teeth to two CW reference lasers at $\lambda=1050$ and 1064~nm. These CW references are themselves pre-stabilized to a common reference cavity with PDH locking. Additionally, the absolute frequency of one of the CW references is measured precisely on an infrared wavemeter with accuracy of $0.3$~GHz.

\begin{figure}[H]
\centering
\includegraphics[width=0.45\textwidth]{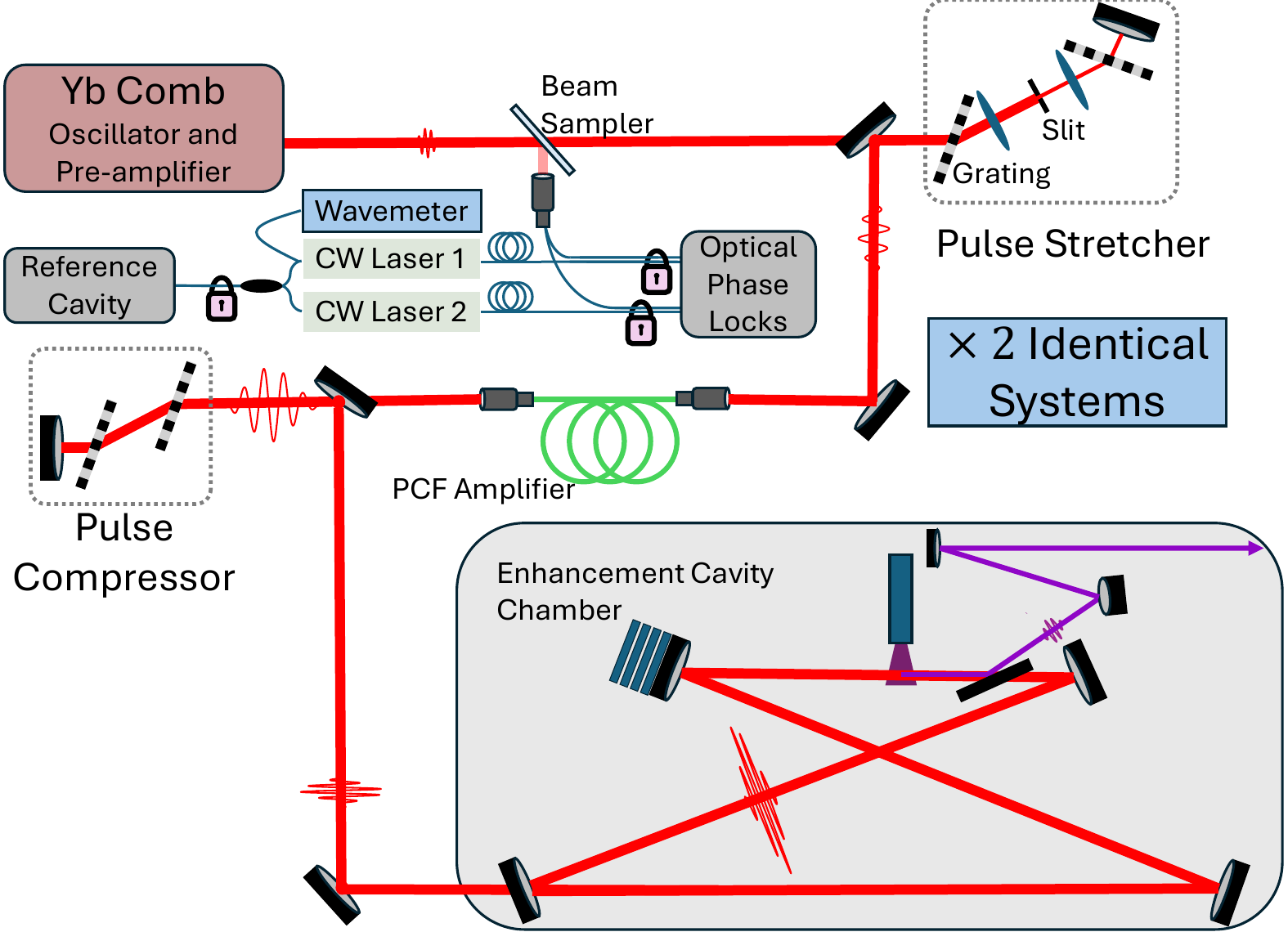}
\caption{\textbf{Ytterbium Comb Schematic.} Illustration of the beampath from an ytterbium fiber oscillator and pre-amplifier, through the pulse stetcher, photonic cyrstal fiber (PCF) amplifier, and pulse compressor before reaching the passive enhancement cavity. Comb modes are controlled with optical phase locks to pre-stabilized continuous-wave (CW) reference lasers, one of which is measured on a wavemeter.}\label{IR_Comb}
\end{figure}

\subsection{Absolute Frequency Calibration}
Because the interference of two optical frequency combs can have a one-to-one mapping from the optical to the RF, we can independently calibrate the absolute optical frequency of our dual-comb measurements. In our current configuration we use two optical locks to stabilize our comb modes. By measuring the comb repetition frequencies, the absolute frequency of one of the CW references with a wavemeter, and the offset phase locks of the combs to this reference, we know the absolute frequencies of all comb modes. Since the high harmonic process generates sum-frequency harmonics from these comb modes, we therefore have precise knowledge of the VUV comb modes as well. By choosing a sufficiently small $\Delta f_{rep}$ ($\sim150$~Hz) each heterodyne interference frequency from these known comb modes maps to a unique RF frequency less than $f_{rep}/2$. This one-to-one correspondence allows us to calibrate the RF-domain measurements directly to the VUV absolute optical frequencies prior to any comparisons with other VUV references or absorption line positions.

\subsection{Dual-Comb Noise Calculations}
As discussed in Section~\ref{sec_discussion}, we calculate expected noise contributions from several sources to understand the limitations in the VUV measurement presented here and to project potential SNR of future measurements with improved devices and more VUV power on detector. Here, we describe the details of these calculations derived from standard DCS theoretical framework~\cite{newbury2010sensitivity}. Considering the DCS noise to arise from four sources---detection noise, shot noise, RIN, and dynamic range---we expect a noise level 
\begin{equation*}
    \sigma=\frac{M\sqrt{\epsilon}}{0.8\sqrt{\tau}}\sqrt{a_{NEP}/{P_c^2}+a_{range}+a_{shot}/{P_c}+a_{RIN}}
\end{equation*}
for $M$ frequency resolved elements, duty cycle $\epsilon=\Delta\nu_{res}/f_{rep}$, averaging time $\tau$, single comb power on detector $P_c$, and noise contributions $a_i$.  

For detection noise, we are limited by stray current noise $I_{rms}\sim1.3~\mu$A through the vacuum chamber feed-through between the PMT and external TIA. This current noise limits the noise equivalent power term to $a_{NEP}=1.7\times10^{-22}$~W$^2$/Hz. The ratio of this stray current noise to the $100~\mu$A maximum output current of the PMT sets the detection dynamic range to be $a_{range}=2\times10^{-10}$~1/Hz. The PMT used has a quantum efficiency at $\lambda=150$~nm of 40\%, so the shot noise contribution is $a_{shot}=1.3\times10^{-17}$~W/Hz. Finally, the VUV laser RIN is measured to be -114~dBc/Hz and fairly constant across the relevant dual-comb carriers, giving the RIN contribution to be $a_{RIN}=1.6\times10^{-11}$~1/Hz.

With $P_c\sim0.6~\mu$W per VUV comb on detector, these terms predict a dual-comb noise limit of $\sigma=0.04$ for $M=1000$ frequency resolved elements with $\tau=68$~s of averaging time, very near the 0.05 VUV noise limit we measure shown in Fig.~\ref{fig_IGM_Averaging}. Additionally, from these terms and the above equation for $\sigma$, it is straightforward to predict the dual-comb noise limit for various VUV powers on detector or given improvements to detection noise or laser RIN mitigation, as shown in Fig.~\ref{fig_NoiseProjection} and surrounding discussion.
\backmatter

\bmhead*{Supplementary information}
Not applicable
\bmhead*{Acknowledgements}
The authors thank Eric R. Hudson and James E. S. Terhune for assistance with the VUV PMT. The authors thank Kelby B. Todd for helpful discussions and gas handling assistance.
\section*{Declarations}
\bmhead*{Funding}
Funding support provided by Air Force Office of Scientific Research [FA9550-20-1-0273]. JJM was funded through the National Defense Science and Engineering Graduate Fellowship Program.
\bmhead*{Conflict of interest/Competing interests }
The authors have no conflict of interest to disclose.
\bmhead*{Ethics approval and consent to participate} 
Not applicable
\bmhead*{Consent for publication}
Not Applicable
\bmhead*{Data availability}
Available upon reasonable request.
\bmhead*{Materials availability}
Not Applicable
\bmhead*{Code availability}
Not Applicable
\bmhead*{Author contribution}

\end{multicols}

\end{document}